\documentclass[aps,pra,twocolumn,amsmath,amssymb,showpacs]{revtex4}            
\usepackage[english]{babel}
\usepackage{latexsym}
\usepackage{graphics}
\usepackage{subfigure}
\usepackage{epsfig}
\begin{document}

\title{Mean field dynamics of superfluid-insulator phase transition
 in a gas of ultra cold atoms}
\author{Jakub Zakrzewski} 
 
\affiliation{
Instytut Fizyki im.Mariana Smoluchowskiego and
Mark Kac Complex Systems Research Center, 
Uniwersytet Jagiello\'nski, ulica Reymonta 4, PL-30-059 Krak\'ow, Poland}
\date{\today}

\pacs{03.75-b,03.75.Kk,03.75.Lm}

\begin{abstract}
A large scale dynamical simulation of the superfluid to Mott insulator
transition in the gas of ultra cold atoms placed in an optical lattice
 is performed using  the time dependent  Gutzwiller mean
field approach. This approximate treatment allows us to take into account most
of the details of the recent experiment [Nature {\bf 415}, 39  (2002)] where
by changing the depth of the lattice potential
an adiabatic transition
from a superfluid  to a Mott insulator state has been reported.
Our simulations reveal a significant excitation
of the system with a transition to insulator in restricted regions of the trap only.
\end{abstract}
\maketitle

\section{Introduction}
                                 
A theoretical suggestion \cite{jaksch} of a possibility to realize one of the standard
models for interacting particles - the Bose-Hubbard (BH) model \cite{fisher,sachdev}  
in a  cold gas placed in an optical lattice has been followed
soon by a seminal  experiment 
\cite{greiner}. The reported realization of a quantum phase
transition between  superfluid (SF) and Mott insulator (MI) phases showed
convincingly that it was possible to control experimentally parameters of the
model practically at will. This triggered several studies involving Bose condensate
\cite{oosten,dicker,menotti,jakmol,damski,boseglass} as well as, more recently 
Fermi-Bose mixtures 
\cite{albus,buechler,lewen} placed on the optical lattices (the reference list must be
not complete bearing in mind that more than 70 papers with ``optical lattice'' in the
title are listed in the cond-mat archive last year only).

At the same time a number of groups \cite{rus,batrouni,zwerger,wessel,roberts}
tried to understand the details of
the very first experiment \cite{greiner} to check the underlying physics.
To imagine the difficulty in modeling the experiment let us recall that
it involves about $10^5$ interacting atoms (bosons) placed in the harmonic 
trap and the three dimensional (3D) lattice potential. 
Such a system is well described by a Bose-Hubbard model with
position dependent chemical potential \cite{jaksch}. Even finding the
ground state of the system for that number of particles and $65\times 65
\times 65$ lattice sites is a formidable task.  
State of the art quantum Monte Carlo (QMC) \cite{rus,batrouni,wessel} 
calculations aimed at
the ground state properties include up to 16 sites in 3D \cite{rus}, 
more sites may be
included in one (1D) or two (2D) dimensional models \cite{batrouni,wessel}.
 These studies, while interesting on their own, can shed little
light on the {\it dynamics} of the system when its parameters are varied.
Except for special exactly solvable models, the efficient simulation of
time-dependent properties of interacting many-body system remains an open
problem although recently quite a progress has been obtained for 1D
systems \cite{vidal,daley}.

It seems, therefore, that the only reasonable and tractable way of
analyzing the dynamics of the discussed experiment is using approximate
methods. To this end we shall use an approach
based on the time dependent variational principle with Gutzwiller 
ansatz. That will allows us to model the details of the 3D experiment \cite{greiner}. The prize for it is similar to that paid in other
approximate treatments - one may always question the extend to which
the approximations allow to describe the properties of the systems studied.
 We hope to convince the reader that the numerical results are at least 
mutually consistent and thus may provide considerable insight into the
dynamics of the experiment.

The discussion of the dynamics is postponed to Section III since
we discuss in the next Section the static mean field solutions for the ground
 state for experimental parameters. Here a comparison with available exact
 QMC results is possible at least. This shall give us
some confidence about the applicability of the mean field approach yielding,
at the same time, the initial state for the dynamics studied later.

\section{Static Mean field for the Bose-Hubbard model}

 The Bose-Hubbard Hamiltonian describing the system takes the form
 \cite{jaksch} 
\begin{equation}
\label{H1}
H=-J\sum_{<i,j>}a^{\dagger}_ia_j +
\frac{U}{2}\sum_{i}n_i(n_i-1)+  \sum_{i}W_i n_i.
\end{equation}
 where $n_i=a^\dagger_i a_i$ is an occupation number operator at site i 
 (with  $a_i$ being the corresponding annihilation bosonic operator), $U$ the
 interaction energy, $J$ the tunneling coefficient and $W_i$ the energy offset
 at site $i$. $\sum_{<i,j>}$ denotes a sum over nearest neighbors.
 Both $J$ and $U$ are functions of the lattice potential and may
 be easily expressed in terms of integrals of the Wannier functions of the lowest
 energy band for cold atoms implementation of the model \cite{jaksch}.
 
 Consider first the standard homogeneous situation in which all $W_i$'s
 are equal. The last term in (\ref{H1}) becomes proportional to the
 (conserved) number of bosons and may be dropped. The only remaining parameter
 of the model is the ratio of $U/J$.  When tunneling dominates the system in its ground state is superfluid while in the
 opposite case it becomes the Mott insulator.
  The borderline between the two
 phases depends on the chemical potential $\mu$. The Mott insulator state is incompressible and is characterized by an integer mean
 occupation of sites. In effect starting from the superfluid at small $U/J$ and a non integer ratio of $N/M$ ($N$ denotes the number of atoms 
 while $M$ the number of sites) and increasing $U/J$ the
 ground state remains superfluid up to highest values of $U/J$ at
 fixed boson density. On the other hand the range of $\mu$ values
 corresponding to the commensurate filling increases with $U/J$. In effect,
 the separation line between a MI and a SF forms characteristic lobes \cite{fisher,sachdev,zwerger}, 
 
   For a detailed discussion
 of the BH model  see \cite{fisher,zwerger}. As we are
 interested in the mean field approximation, let us just 
 quote Zwerger \cite{zwerger} saying "In two and three-dimensional lattices,
 the critical value for the transition from a MI to a SF is reasonably well described by a mean-field approximation". In one dimension the
 mean field approximation is much worse \cite{zwerger}.
 
In the presence of the additional potential, e.g. the harmonic trap, local energies $W_i$  depend on the sites location. Then the effective chemical potential at each site becomes $\mu_i=\mu-W_i$. As pointed
out already in \cite{jaksch} this will lead for large $U/J$ to a
shell like structure with MI phases with different integer occupations
(highest in the middle assuming attractive binding additional potential) separated by SF regions. This picture has been nicely confirmed in quantum Monte Carlo calculations both in 1D \cite{batrouni} and in 3D \cite{rus}.

The latter exact results are of particular interest for us since they
allow for a comparison with the mean field approximation. In 
\cite{rus}  a 3D  $16\times 16 \times 16 $ lattice is considered
with different values of  $U$ and $J$ parameters as well as the harmonic trap.
To find the mean field ground state we minimize 
 \begin{equation}
\label{static}
<E>=<G|H- \mu {\hat N}|G>,
\end{equation}
where ${ \hat N}=\sum_i n_i$
and  $|G>$  is the Gutzwiller trial function
\begin{equation}
\label{gut}
|G>= \prod_{i=1}^{M}(\sum_{n=0}^{n_m}
f_n^{(i)}|n>_i).
\end{equation}
The number of  parameters $f_n^{(i)}$ depends on the number of
sites (here $16^3$) as well as the maximal occupation at a given
site $n_m$. The 
average maximal occupation at the center of the trap is $2$
fot the data considered in \cite{rus}. Therefore, it is
sufficient to take $n_m=7$. That yields a minimization procedure 
over  32768 parameters. Such a number of parameters must lead to 
a spurious local minima, unless a very good estimate exists for the
initial set of $f_n^{(i)}$'s (i.e. the initial $|G>$). Fortunately
such a guess is quite obvious and is often termed a local mean field
approximation. Namely at each site $i$ one takes a solution for 
 $f_n^{(i)}$'s corresponding to the homogeneous BH model with 
 the effective chemical potential $\mu_i=\mu-W_i$. Provided $W_i$
 changes smoothly from site to site, such an approach should be an excellent approximation to a full
 mean field solution. And indeed it is,  we have
 found for the data discussed below that the initial and final $<E>$ [see Eq.~(\ref{static})] differ by
 at most 2\%; the number of iterations of  standard Numerical Recipes minimization packages \cite{NR} is slightly bigger than the number of parameters.

 \begin{figure} 
\begin{center}
\psfig{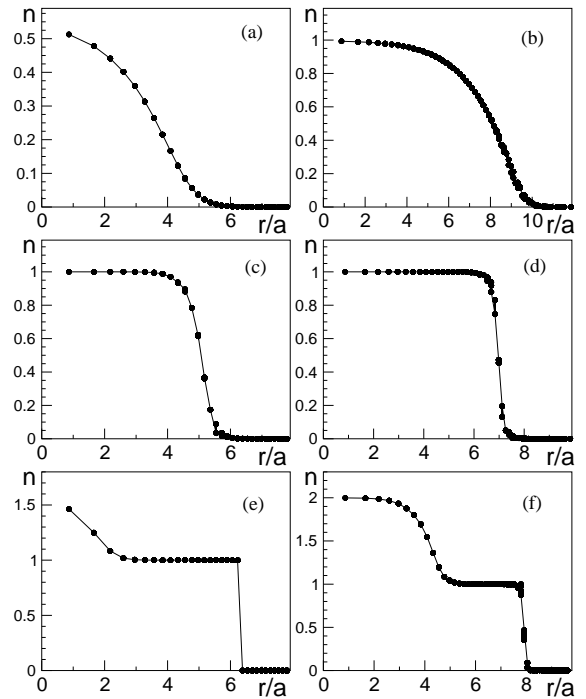}
\end{center} 
\caption{Mean field particle density distribution (on-site filling factors) as a function of the distance from the center of the trap measured in the units of the lattice constant $a$. Filled circles correspond to numerical results, lines are drawn to guide the eye.
The parameters of the BH model match  those quoted in Fig.~1 of
\protect\cite{rus}, see the text for numerical values.
 Comparison with the latter results obtained within ``exact'' 
 quantum Monte Carlo reveals the accuracy of the mean field
approximation.}
\label{svist1}  
\end{figure}

The results obtained are presented in Fig.~\ref{svist1} in the same
form as the corresponding plot in \cite{rus} to make the comparison
easier.
The values of parameters correspond to those taken in \cite{rus}. The
notation used by us differs slightly from that of \cite{rus}, in particular
the tunneling constant $J$ is denoted as $t$ in \cite{rus}. For the purpose of this figure all parameters are expressed in units of $J$, i.e. $J=1$.
To make the comparison with notation used in \cite{rus} easier,
the coefficient in the term proportional to the number of atoms at a given site
in (\ref{static}) i.e. the difference of the on-site energy
$W_i$ and the chemical potential $\mu$ 
is expressed as $W_i-\mu=-U_0+ U/2+\kappa {\bf x}_i^2$, where ${\bf x}_i$ is a 
position vector of  site $i$ with ${\bf x}_i^2$ being the square of the distance
of the $i$-th site from the center of the harmonic trap measured in units of the
lattice constant. With such a definition the parameters used in Fig.~\ref{svist1} take the values: panel 
(a): $U=24.$, $U_0=-11.08$, $\kappa=0.19531$;
(b): $U=32.$, $U_0=-28.08$, $\kappa=0.19531$;
(c): $U=80.$, $U_0=-65.0$, $\kappa=0.97656$;
(a): $U=80.$, $U_0=-90.0$, $\kappa=1.03062$;
(a): $U=80.$, $U_0=-120.08$, $\kappa=2.00375$; and
(a): $U=80.$, $U_0=-150.0$, $\kappa=1.75781$.
A comparison of Fig.~\ref{svist1} with Fig.~1 of \cite{rus} indicates that, as far as the average occupation at different sites is concerned, the mean field solution is
in excellent agreement with the quantum Monte Carlo results.

 \begin{figure*}[ht] 
\begin{center}
\psfig{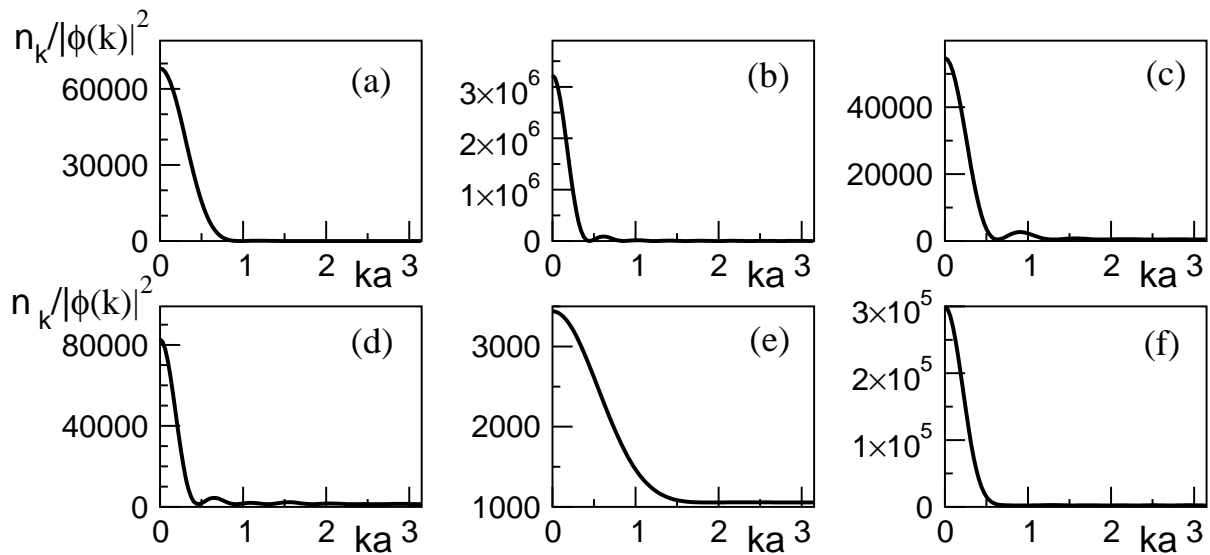}
\end{center} 
\caption{The rescaled mean field momentum distribution (in arbitrary units) in the first Brillouin zone in the (0,0,1) direction derived
from (\ref{mom}) for data presented in Fig.~\protect\ref{svist1}.
The distributions are to be compared with the corresponding ``exact''
distributions  in Fig.~2 of \protect\cite{rus}. Observe that mean field
result differs from the exact distribution for the case (d) only --
see discussion in the text. 
}
\label{svist2}  
\end{figure*}

Instead of the occupation at different sites one may take a look
at the momentum distribution, i.e. the quantity closely related to that  measured
in the experiment (see \cite{greiner} and the discussion below).
 The momentum distribution is given by \cite{rus}
\begin{equation}
\label{mom}
n_{\mathbf k}= |\phi ({\mathbf k})|^2\sum_{i,j}
e^{i{\mathbf k}\cdot({\mathbf r}_i- {\mathbf r}_j)}<a^\dagger_ia_j>,
\end{equation}
where ${\mathbf k}$ is the wavevector, $\phi ({\mathbf k})$  a Fourier transform of the Wannier site-function. The latter yields a broad bell-shaped background and provides merely information about the lattice. The relevant information about the atoms is contained
in the Fourier transform of  $<a^\dagger_ia_j>$. In the mean field
approximation this factorizes for different $i,j$ $<a^\dagger_ia_j>\approx<a^\dagger_i><a_j>$. Such a factorization
seems quite drastic and one may expect significant differences between 
the momentum distributions obtained from QMC and within the mean field approximation. It is really not so, however, for bosons
in a harmonic trap as visualized in Fig.~\ref{svist2} for the mean
field. That figure
should be compared with QMC results presented in Fig.~2 of \cite{rus}. Observe that differences appear only for panel (d), the
exact results yield significantly broader momentum distribution. As
discussed in \cite{rus} almost no SF fraction is present in the QMC result corresponding to panel (d). Then the factorization must
affect strongly the momentum distribution since in a vast majority of sites $<a_i>=0$. Clearly, however, as long as some SF fraction is
present in the system the mean field momentum distribution closely resembles the exact quantum results.

This apparent quite close agreement between QMC
results and the mean field approximation for $16\times 16\times 16$
lattice and about $10^3$ atoms is very encouraging in view of
realistic experimental conditions \cite{greiner}. Here both the external potential
changes less rapidly (the size of the lattice is now $65\times 65\times 65$) and number of atoms exceeds $10^5$ thus one may expect that the 
mean field approximation works even better.

While the test described above have been taken for some chosen 
(by authors of \cite{rus}) values
of $U$, $J$, as well as the trap frequency, to simulate the experiment
we have to determine first the relevant range of parameters. From now on we shall measure the quantities of dimension of energy in the units of the recoil energy of $^{87}$Rb atoms for light with a wavelength 
$\lambda=2\pi/K= 852 {\mathrm nm}$, i.e. $E_r=\hbar^2K^2/2M$, where
$M$ is the mass of $^{87}$Rb atoms. The depth of the optical lattice 
$V$ changes from 0 to $22E_r$. Finding Wannier functions for
different values of $V$ \cite{bodzio} we evaluate the corresponding 
$U(V)$ and $J(V)$ values. The energy offset at each site
 $W_i$ has two components in the experiment \cite{greiner}. One is the harmonic  magnetic trap potential
 (time-independent), another is due to the Gaussian intensity 
 profiles of lattice creating laser beams. The latter may be also approximated by a
 harmonic term \cite{greiner} the corresponding frequency is then
 dependent on $V$.
 
  \begin{figure*}[ht] 
\begin{center}
\psfig{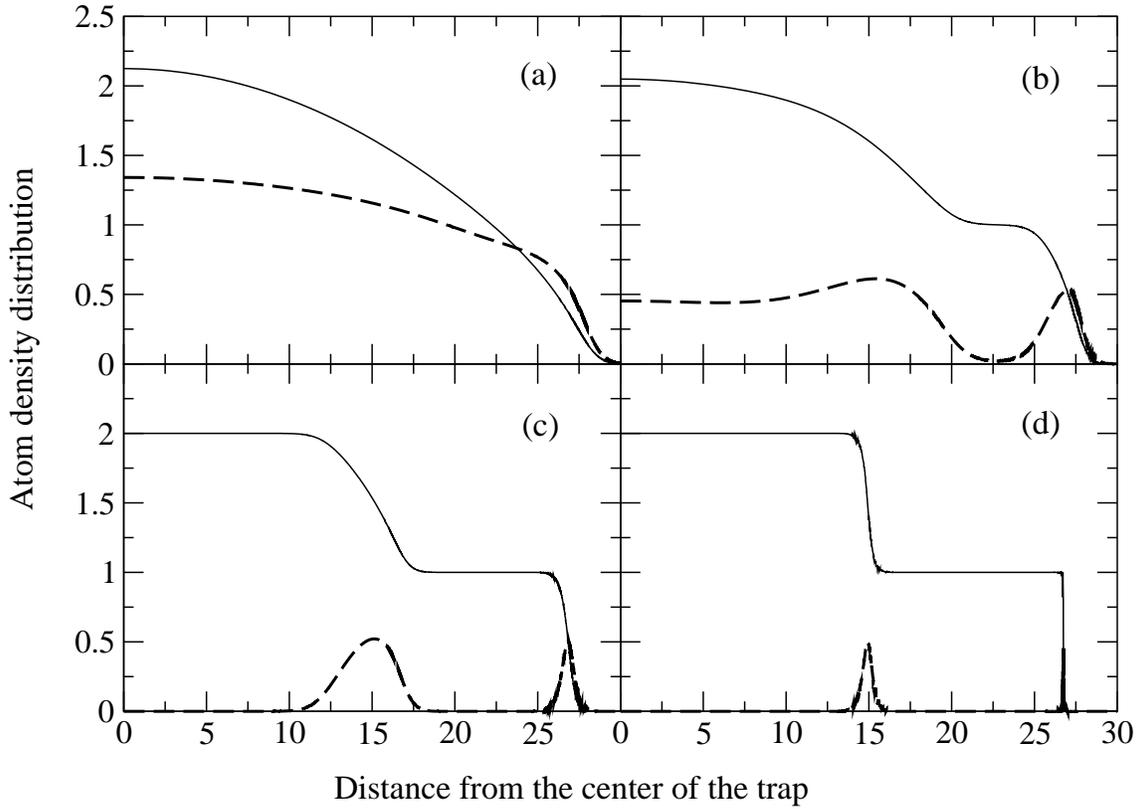}
\end{center} 
\caption{Solid lines represent mean field atom density distribution (on-site filling factor - $n$) as a function of the distance from the center of the trap
(measured in units of the lattice constant). Dashed lines represent the  corresponding variances at different sites
$2\sigma_i^2=2(<n_i^2>-<n_i>^2)$.
 The total number of atoms and the SF factor
$\gamma_{SF}$ (see text) for each plot are (a): 
$V=9E_r$, $N=99771$, $\gamma_{SF}=0.95$; (b):
$V=13E_r$, $N=99502$, $\gamma_{SF}=0.40$; (c):
$V=16E_r$, $N=95408$, $\gamma_{SF}=0.11$; (d):
$V=22E_r$, $N=94172$, $\gamma_{SF}=0.01$. Small indents seen in the lines
 in panels (c) and (d) are due to anisotropy induced by a cubic lattice in
an isotropic (harmonic) trap, for further details see text.
}
\label{staticf}  
\end{figure*}
 
 To find the mean field ground state for different $V$
 values and the number of atoms of the order of $10^5$  one needs to 
 solve a minimization problem over $2\times 10^6$
 parameters (with $n_m=7$
 as before)
 which is hardly manageable. One may, however, use the symmetry of the problem (cubic lattice combined with spherically 
 symmetric trap) to significantly reduce that number. Let $i,j,k$
 count  the sites in $x,y,z$ directions, respectively with each
 index taking the values from $-32$ to $n_s=32$ (yielding 65 sites in each direction). Due to the ground state symmetry it is enough to
 consider only the sites with $0\le i\le j\le k\le n_s$ which 
 reduces the number of minimized parameters to about 48 thousands.
 Needless to say we have checked on the smaller $16 \times 16 \times 16$ problem that the symmetry reduced problem yields the same ground state as the full minimization.

 The results obtained are presented in Fig.~\ref {staticf} and are
 practically indistinguishable from the initial guess i.e. a wavefunction coming from local mean field approximation discussed above. The chemical potential $\mu$ has been  adjusted (for each case) to have the average number of atoms 
 $N=<{ \hat N}>=\sum_i <n_i>$ 
 around $10^5$. This leads to more than two atoms (on average)
 per site in the center of the trap.
  To characterize whether the
 state is closer to being superfluid or Mott insulator we define
  the superfluid factor $\gamma_{SF} =1/N \sum <a_i><a^{\dagger}_i>$.
 This factor is zero for pure MI state (when $<a_i>=0$ as each node is in a Fock
 state) and reaches unity for Poissonian statistics at each node. While
 obviously it is not a ``proper'' order parameter for the phase transition,
 it seems to be convenient for characterizing the states obtained.
 Using this
 factor we can quantify states shown in Fig.~1, noting first a general
 qualitative agreement with experimental findings \cite{greiner}.
 The case $V=9E_r$
 seems almost fully superfluid (with, however, strongly subpoissonian 
 statistics \cite{orzel} at each site), the case $V=13E_r$
 shows first traces of insulator phase (integer occupation of sites with vanishing
 variance, the transition is completed for significant fraction of sites at $V=16E_r$
 while for the deepest lattice $V=22E_r$ SF fraction is restricted to very narrow
 regions separating different integer occupations. A careful reader may notice
 small indents visible in lines in panels (c) and (d) of Fig.~\ref{staticf}. They are due to anisotropy induced by a cubic lattice in an isotropic harmonic trap. The data, presented necessarily as a function of the radial distance
 contain occupations along the axes, the diagonals, and other not equivalent directions in the cubic lattice. The smallness of the indents indicates that the symmetry of the trap is dominant. 
 
 \section{Time-dependent mean field dynamics}
   
The results obtained for mean field ground state in realistic situations, shown in the previous Section, seem quite encouraging. Yet, in themselves they can say little about the dynamics of the system when
the lattice depth $V$ is varied. In an attempt to address this important issue we shall use a time-dependent version of the mean field
approximation. To this end we employ the time-dependent variational
principle \cite{jakmol}  looking for the minimum of

\begin{equation}
\label{var}
<G(t)|i\hbar \frac{\partial}{\partial t} -H(t)+ \mu {\hat N}
|G(t)>,
\end{equation}
with $H(t)$ being now the time dependent Hamiltonian. The time dependence is implicit via the dependence of the BH Hamiltonian $H$ 
 on $U,$ $J$ and $W_i$, that in turn depend on $V$. The chemical potential
$\mu$  becomes also time dependent when system parameters
are varied. $|G(t)>$,  the variational  wavefunction, is assumed
 in the standard Gutzwiller-type form (\ref{gut}),
 with $f_n^{(i)}(t)$ now being time-dependent.The very
same approach has been successfully 
applied recently to the formation of molecules \cite{jakmol,damski},
the treatment of the
disordered optical lattices \cite{boseglass} as well as for
determining the phase diagram in Bose-Fermi mixtures \cite{lewen}.

The minimization of (\ref{var}) yields the set of first order
differential equations for $f_n^{(i)}(t)$ :

\begin{eqnarray}
\label{T}
i\frac{d}{dt}{f}^{(i)}_n &=& \left[\frac{U}{2}n(n-1)+n (W_i-\mu)\right]f^{(i)}_n-
\nonumber \\
&&J\left[\Phi^{\star}_i \sqrt{n+1} f^{(i)}_{n+1}+ \Phi_{i} \sqrt{n} f^{(i)}_{n-1}\right] ,
\label{dyn}
\end{eqnarray}
where
$\Phi_{i}=\sum_{<j>}<G(t)|a_j|G(t)>$ (the sum, as indicated by subscript in
brackets is over the nearest neighbors only). The nice feature of the evolution resulting from equations (\ref{T}) is that the average number
of particles $N=<\hat N>$ is an exact constant of the motion \cite{jakmol}.
Naturally when the parameters of the BH model, e.g. $U$ and $J$, change
the chemical potential corresponding to the mean field solution with a
given number of particles $N$ also changes. The dynamics of $\mu$ cannot
be obtained from (\ref{T}) only. One can find it, however, following the evolution of two states $|G_1>$ and $|G_2>$ with slightly different average number of particles $N_2=<G_2|\hat N|G_2>=N_1+\delta N= <G_1|\hat N|G_1> 
+\delta N$. The chemical potential at given $t$ may be then approximated by
$\mu(t)=(<G_2(t)|H(t)|G_2(t)>-<G_1(t)|H(t)|G_1(t)>)/\delta N$ and adjusted at
each time step \cite{lewen}. This is the approach used in the numerical results presented below.

 \begin{figure}[ht]
\begin{center}
\psfig{file=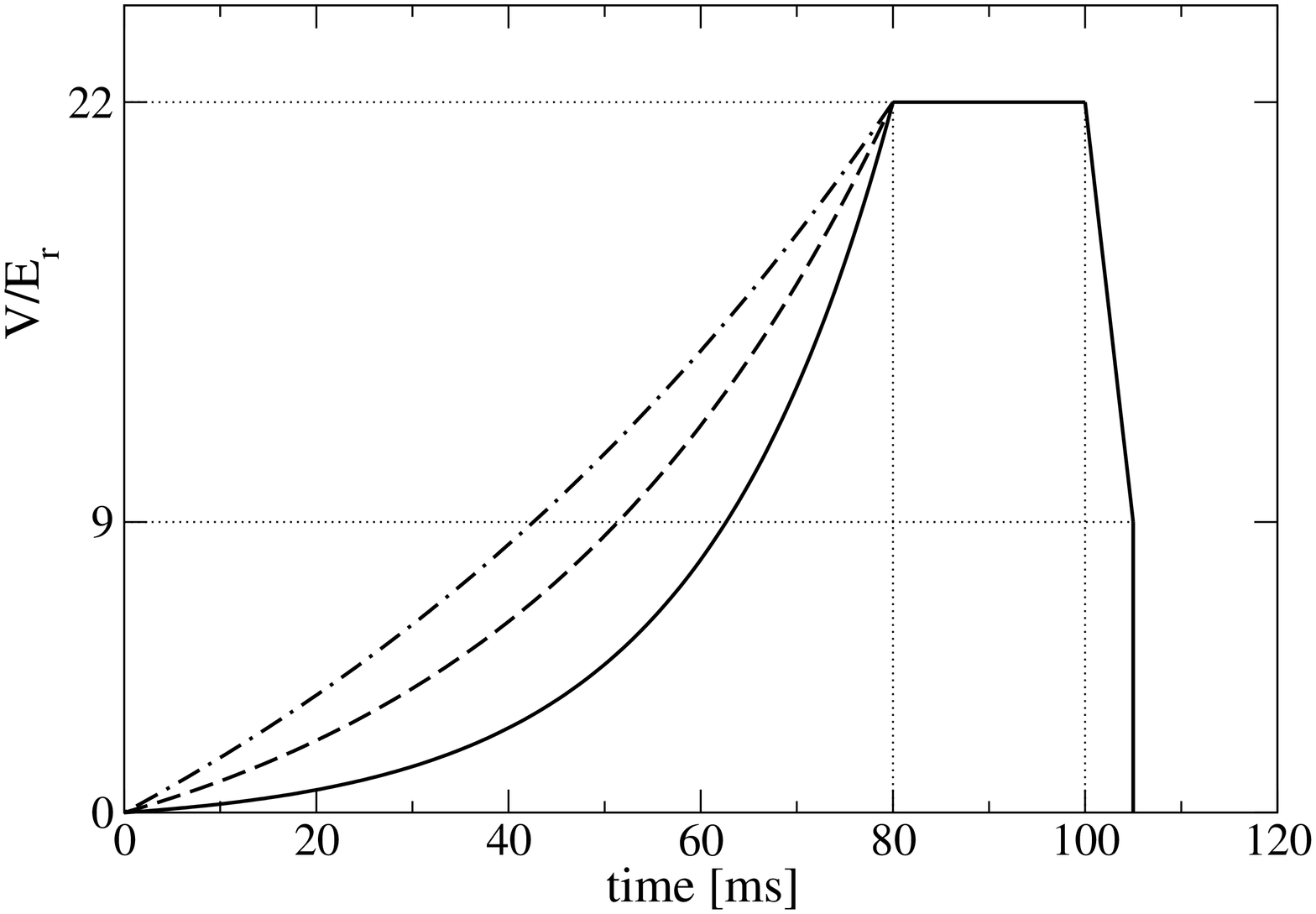,width=7.cm,angle=-90}
\end{center} 
\caption{The experimental time profile (solid line) of the lattice potential depth $V$
measured in the recoil energy $E_r$ units. The initial exponential increase
with a time constant $\tau=20$\ ms
is followed by a flat part and a subsequent decrease to $V_f=9E_r$ with a linear ramp with varying slope. Dashed line corresponds to the exponential time constant $\tau=40$\ ms, dash-dotted $\tau=$\ 80 ms. Thin dotted lines indicate
particularly interesting values of time and $V$ -- see text for discussion.}
\label{timescan}  
\end{figure}

Since we want to follow as closely as possible the experiment
\cite{greiner} let us recall its main features.
The experiment has three stages after loading the
harmonic trap with Rb condensate - compare Fig.~\ref{timescan}. Firstly, the optical
lattice depth $V(t)$ was increased in 80 ms (using exponential ramp with time constant $\tau=20$\ ms) from  the initial zero value (when the harmonic trap was present only)
to $V_{max}=22E_r$, where $E_r$ is the recoil energy of Rb atoms. The sample was
then held for 20ms at $V_{max}$. Finally $V(t)$ was decreased with the linear
ramp to $V_f=9E_r$ with different speed. At any stage the experiment could be
interrupted by rapidly switching off all laser beams building up the lattice as well
as the magnetic trap. The freely expanding atomic cloud, after some delay, was
recorded by a destructive absorption  imaging, yielding the signal which
reflects the momentum distribution \cite{zwerger,rus}. Since the absorption
images are taken along two orthogonal axes the quantity measured is in fact
the integrated momentum distribution \cite{rus}:
\begin{equation}
N(k_x,k_y)\propto \int dk_z n_{\mathbf k}.
\label{average}
\end{equation}

For clouds released from
low optical lattices when tunneling dominates and the superfluid behavior is
expected the signal reflects Bragg peaks due to interferences of atoms coming
from different lattice sites. At increased lattice depths  above
$13E_r$ the interference maxima become immersed in an incoherent background 
disappearing practically at $20E_r$. This behavior was associated with the
quantum phase transition from SF to MI phase \cite{greiner}. Most interestingly
the coherence of the sample may be rapidly recovered when the lattice depth is
decreased (third stage of the experiment) as measured by the width of the
central interference peak which decreases almost to its original value at
$V=9E_r$ in about 4ms.
 
\begin{figure*}[ht] 
\begin{center}
\psfig{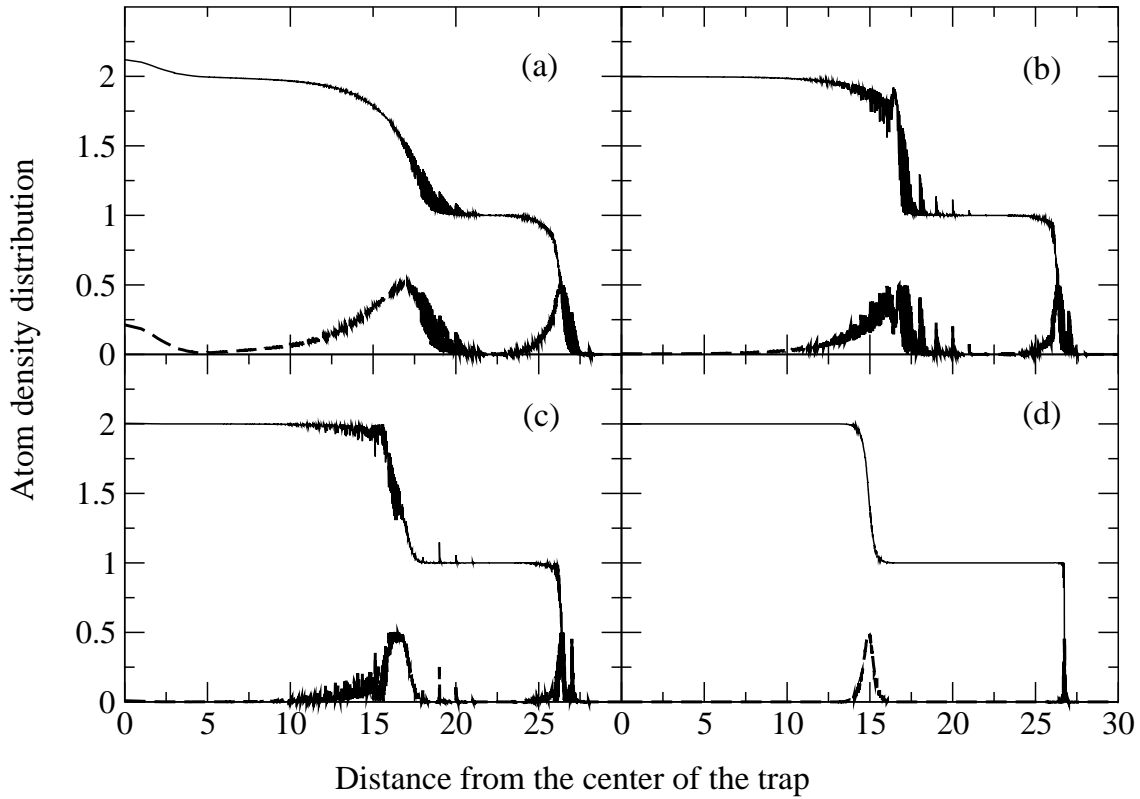}
\end{center} 
\caption{Atom density distribution (on-site filling factor) after a mean field evolution starting from the mean field ground state at $V=9E_r$ at the final value of $V=22E_r$ for different $V(t)$ time dependence. (a) corresponds to
the exponential time scale of 20 ms, (b) to 40 ms, (c) to 80 ms. The longer the time scale, the slower the change of $V(t)$ in the considered range as
can be seen in Fig.~\protect\ref{timescan}. Thick dashed lines in (a) to (c)
present twice the  variance of the on-site occupation showing that the insulator regions in (b) and (c) are much larger than in (a). Panel (d)
repeats, after Fig.~\protect\ref{staticf}, the mean field ground state at $V=22E_r$ for comparison. Observe that indents in the lines for dynamically 
evolved wavefunctions are much more pronounced than in the static (d) case. As before these indents are due to different symmetries of the cubic lattice and sperically harmonic isotropic trap. The conflicting symmetries lead to anisotropic excitations in the dynamical situations when the potential parameters are varied. Those excitations are responsible for the indents.}
\label{dyna}  
\end{figure*}

In the former applications of the time-dependent mean field approach \cite{jakmol,damski,boseglass,lewen} the time-dependence of system's
parameters was assumed to be sufficiently slow to assure adiabaticity. In
effect the mean field ground state has been followed by applying the
time-dependent equations for   $f_n^{(i)}(t)$'s (\ref{T}). Here we
have a similar situation since it is claimed \cite{greiner} that the changes
in time of $V(t)$ are made sufficiently slow to keep the system in the many-body ground state. Having the mean field ground states for different $V$ values we can (within the mean approximation) test this adiabaticity assumption.

Looking again at the time profile depicted in Fig.~\ref{timescan} one may notice that the ramp used in the experiment leads indeed to a very slow increase of $V(t)$ initially, however changes of $V(t)$ become relatively
rapid about and above $V=9E_r$, i.e. in the region where the transition from SF to MI is supposed to take place. Taking as the initial state the mean field ground state at $V=9E_r$, our simulations show 
that to assure adiabaticity a small change of $V$ on the superfluid side (say from $V=9E_r$ to
$V=9.1E_r$)  requires about 20ms (one needs 40 ms
for a loop from $V=9E_r$ to $V=9.1E_r$ and back to keep the overlap on the initial state of 
the order of 99 percent). That strongly indicates that a much longer time is needed to traverse adiabatically  the whole interesting region from $V=9E_r$ to $V=22E_r$. And that change is realized in about 20ms in the experiment.

To test the adiabatic issue further  we shall concentrate in the following
on the regime  above $V=9E_r$ containing the quantum phase transition.
Starting again from the Gutzwiller mean field ground state at $V=9E_r$ we simulate the
time evolution up to $V=22E_r$ (with experimental time profile). We may compare the dynamically obtained wavefunction plotted in Fig.~\ref{dyna}(a) with the
mean field ground state at $V=22E_r$ (bottom right panel (d) in the figure).  While the ground state
has an insulator character almost everywhere in the trap  with
$\gamma_{SF}=0.01$, the dynamically
evolved wavefunction, by comparison, 
seems to reflect an excited wavepacket and it has
rather small regions where the occupation of sites is close to
integer with vanishing number variance.
 The corresponding $\gamma_{SF}=0.12$ confirms the presence of a relatively large superfluid region.

To show that the effect is really due to the too fast increase of the lattice
depth we have modified the experimental time profile slightly, by changing the
exponential constant $\tau$ from 20 ms to 40ms (or 80 ms). That makes the initial
rise of the laser intensity (and the lattice depth) more uniform in time - 
compare Fig.~\ref{timescan}. Observe that while the full duration of the 
first stage remains the same, the interval of time spend on the increase of
the lattice depth from $V=9E_r$ to $22E_r$ increases from below 20 ms (experimental profile), to about 30 ms (for $\tau=40$\ ms) or about 37 ms
(for $\tau=80$\ ms). Starting again from the mean field ground state at $V=9E_r$ we obtain the atom density profiles shown in Fig.~\ref{dyna}(b)
and Fig.~\ref{dyna}(c), respectively. Observe that the regions of insulator
behavior for both cases are much larger than observed previously. The
corresponding superfluid factors are $\gamma_{SF}=0.062$ for $\tau=40 ms$
and $\gamma_{SF}=0.048$ for $\tau=80 ms$. While the final distributions still show signs of significant excitations, the insulator character becomes dominant for (c) and also for (b) case.

Keeping the overall duration of the laser intensity turn-on at 80 ms and enlarging
the final stage comes at the price that the initial rise from $V=0E_r$
to $V=9E_r$, very smooth in the experiment \cite{greiner} for $\tau=20$\ ms,
becomes sharper for larger $\tau$ (compare Fig.~\ref{timescan}). Thus larger
$\tau$ may lead to some excitation at the initial creation of the lattice, not
apparent in our simulations since we start from the ground state at $V=9E_r$.
Trying to keep the total duration of the experiment as short as possible
(to avoid, for example, the decoherence) one can still imagine a slightly more
sophisticated pulse rise with say $\tau=20$\ ms initially up to say $V=9E_r$
and further increase with a larger time constant, say 40 ms. While the duration
of the experiment increases by 10\%  only,
 the degree of the excitation of the final wavepacket becomes much smaller and the insulator character much more pronounced.

 In the experiment \cite{greiner} the atomic density distribution is not 
 measured directly. The presence of the Mott insulator layer has been detected by observing  the resonance in the excitation spectrum around the interaction energy $U$. Clearly the size of the corresponding peak is  
 related to the number of atoms in the insulator layers. Our results indicate
 that an appropriate suggested change in the time profile of $V(t)$ should
 increase the size of insulator regions and thus enhance the resonant
 peak in the excitation spectrum.
 
Let us mention also that qualitatively similar results to that depicted in Fig.~\ref{dyna} are obtained starting the evolution from different value of $V$
say $V=8E_r$ as long as the initial $V$ is sufficiently big but smaller than
the beginning of the SF-MI transition regime (around $V=13E_r$).
 In the experiment the lattice depth is increased from the initial zero value
 but for that case the Bose-Hubbard Hamiltonian (\ref{H1}) is not a good model to describe the cold atoms physics. This discrete lattice model is appropriate \cite{jaksch,zwerger} when the atoms have to remain in the lowest vibrational state at each site that is for a sufficiently deep lattice.
 
\begin{figure*}[ht] 
\begin{center}
\psfig{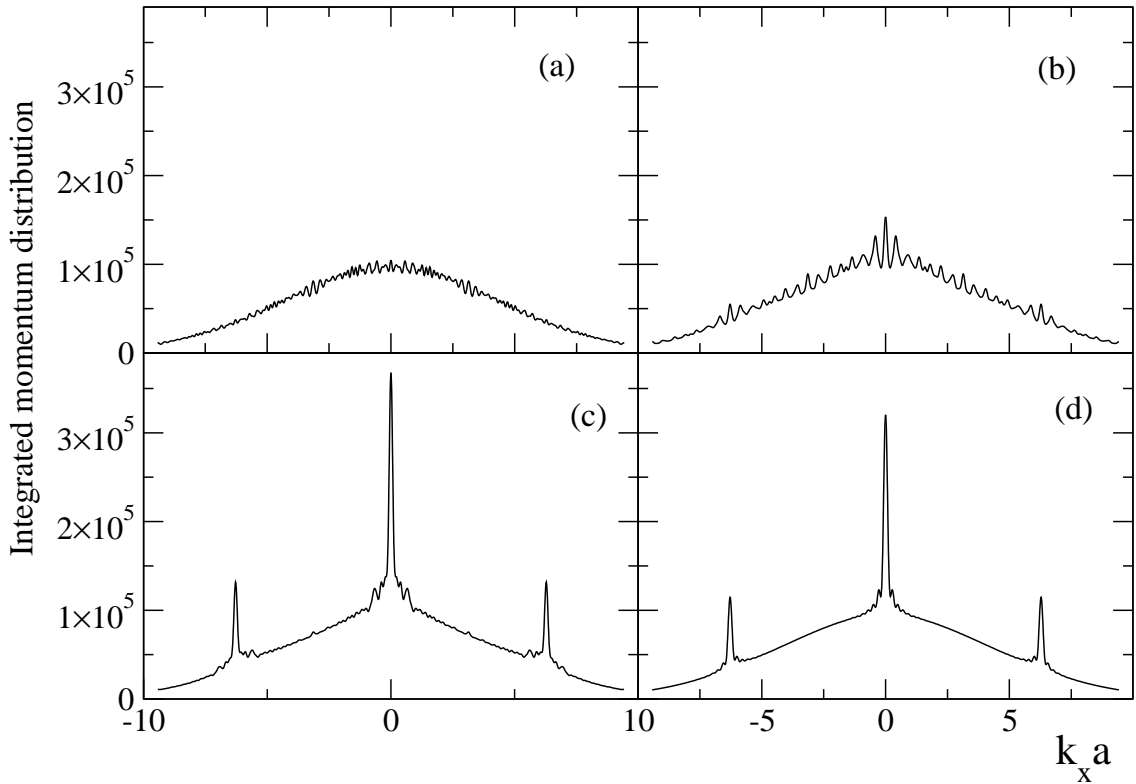}
\end{center} 
\caption{Integrated momentum distributions (\ref{average}) 
for states shown in Fig.~\ref{dyna} (panels match one another).
 The cut $N(k_x,0)$ along $k_y=0$ is shown only. 
 Observe that for the time profile of the experiment no narrow peak is visible. in agreement with the experiment.
For more adiabatic time scale -- panel (c), as well as for the ground state --
 (d) -- the narrow
peak in the momentum distribution is clearly visible. For a further discussion
see text.
}
\label{dynaf}  
\end{figure*}

 It is most interesting to compare the integrated momentum distributions (\ref{average}) 
 corresponding
 to the position distributions shown in Fig.~\ref{dyna}. The cuts along  
$k_y=0$ are presented in Fig.~\ref{dynaf}. In contrast to Fig.~\ref{svist2} we include now the
Fourier transform of the Wannier function [compare eq.~\ref{mom})] so the quantity plotted 
matches the experiment (for a better quantitative picture, 
we prefer the cut along $k_y=0$ instead of the three-dimensional color plot
used oroginally to represent the experimental data).   
 Panel (a) corresponds to the experimental profile. 
 Observe that the integrated momentum distribution consists of a broad peak
in a very nice agreement with the experiment \cite{greiner}. Note, however,
that for other, ``more adiabatic''  time profiles the narrow central structure
emerges [compare (b)] becoming quite pronounced both for (c) as well as
 for
the mean field ground state momentum distribution -- depicted in panel (d).
The absence of the narrow Bragg peak in the experiment seems to be, therefore,
not related to the transition to insulator phase as suggested in \cite{greiner}. 
The persistence of the narrow peaks in Mott regime has been already noted in \cite{rus} 
in a model  static quantum Monte Carlo study on a smaller lattice. 
Result presented in (d)
indicates that the conclusion of the authors \cite{rus}
 concerning the ground state momentum distribution extends also
   to a realistic sample.
The dynamical results presented in Fug.~\ref{dynaf} suggests that fading of the Bragg peaks
and the appearance of the broad integrated momentum distribution can be 
associated with a significant excitation of the sample rather than the
transition to Mott insulator regime. 

The reliability of the dynamical mean field approach may be tested further
in an attempt to
reproduce ``restoration of coherence'' part of the experiment \cite{greiner}.
In the experiment, after reaching $V=22E_r$ the lattice depth is kept constant 
for 20ms and then decreased back to $V=9E_r$ with a linear ramp of duration$t$.
 It is shown that the time 
needed to  restore the narrow interference pattern in the integrated momentum distribution
is of the order of 4ms. This phenomenon is associated with restoring the coherence in the
sample. 

This {\it interpretation} is in contradiction with results already presented in Fig.~\ref{dynaf}
-- the existence (or the lack of it) of the narrow peaks
 seems not to be solely related to the  coherence of the sample but also to
its excitation.
The question, however, remains whether the dynamical mean field calculation can reproduce
the {\it results} of the experiment, i.e. the dependence of the width of the interference 
pattern on the time duration of lowering the lattice.

\begin{figure}
\psfig{file=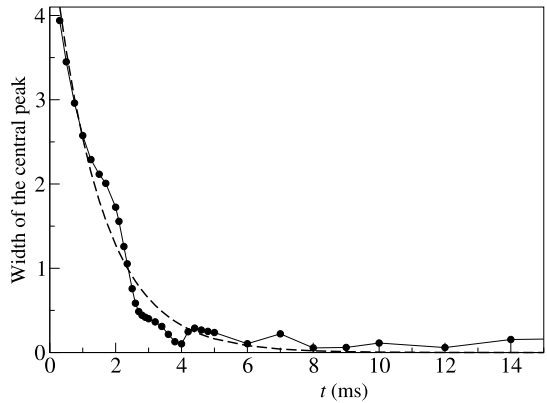,width=8.5cm}
\caption{Half-width of the central interference peak for different ramp down times $t_r$ obtained by a lorentzian fit of the integrated momentum distribution -- compare (\protect\ref{average}).
 Filled circles are connected by a line
to guide the eye. Dashed line is a single exponential decay with a time constant
$\tau=1.45$ ms.}
\label{figfin}  
\end{figure}

To answer this question  we make a 
simulation (the results are shown in Fig.~\ref{figfin}),
 starting from the static solution in the SF regime (taken for the convenience
at $V=9E_r$ again) increasing exponentially the lattice height as in the experiment
\cite{greiner}, the subsequent 
delay of 20 ms at $V=22E_r$ and a linear ramp-down with various slopes. Note that 
 the shape of the curve as well as the time
scale of restoring the narrow interference pattern is in quite good agreement with the experiment.
 While
the experimental data could be fit with a double exponential decay with two time
scales, our mean field data are reasonably reproduced with a single exponential decay
with time scale $\tau=1.45$ ms. This nicely corresponds with the shorter time scale
of the experiment (0.94 ms). The obtained  time scale  is also of the order of a typical single
 tunneling time ($1/J$ in appropriate units) to the nearby site. Our  numerical data fail to
 reproduce the second experimental time scale. This points out to the possibility that 
 it can be associated indeed with a long range correlation between sites. Such a a long
 range correlation
 should not manifest itself in our mean field simulations -- the
 Gutzwiller wavefunction neglects entanglement between sites.

The observed quite good agreement of the obtained widths of the momentum distribution with the experiment \cite{greiner} seems to be quite 
a spectacular success of the dynamical mean field simulation bearing in mind its
simplicity. The fact that the mean field approach works so well may be, in our opinion, attributed to the fact that the dynamics takes place in the regime where superfluid fraction remains significant. Then the mean
atomic field $\Phi_i$ does not vanish allowing for semiclassical (mean-field)
description.
 Our results suggest 
 that the system has quite a long memory and remembers
  that it was originally 
a SF. This fits nicely with the excited wavepacket-like character of dynamically
obtained  wavefunction clearly visible in Fig.~\ref{dyna}(a).

\section{Conclusions}

To summarize, it has been shown that the mean field Gutzwiller approximation allows one
to simulate a dynamics of inhomogeneous Bose-Hubbard model taking into account
realistic experimental conditions. The accuracy of the approximation cannot be
controlled which is the major drawback of the present approach (a comparison with
exact dynamics for small systems will lead us nowhere since then the mean field
approach is known to fail). On the other hand a comparison with the available data 
seems quite encouraging. Accepting mean field predictions we may confirm that indeed
the transition from superfluid to Mott insulator takes place (albeit in a small part of the 
sample) in the experiment 
\cite{greiner}. On the other hand the claim that the first stage of the experiment
  is performed adiabatically
assuring that the system remains in its many body ground state (and thus a genuine
textbook quantum phase transition \cite{sachdev} is realized) seems questionable.
We confirm the suggestion of \cite{rus} that fading of narrow peaks in the momentum
distribution should not be assotiated with the transition to the Mott insulator regime.
Rather it is a dynamical effect.

We suggest that optimization of the lattice depth time dependence (i.e.
laser intensity profile) may help to enlarge the insulator regions making
the transition more adiabatic. That
may be detected by measuring the size of the peak in the excitation spectrum
of the system.

Lastly, let us mention, that a very recent preprint \cite{jaksch04} reports 
a study of exact dynamics
of the model using the method  of \cite{vidal,daley}. However, the results consider
at most 49 atoms in 40 sites of 1D lattice.

Participation of D. Delande at the early stage of this work is appreciated as well as
discussions with B. Damski, Lewenstein and K. Sacha.
This work was
supported by Polish Committee for Scientific Research Grant Quantum Information and
Quantum Engineering,  
PBZ-MIN-008/P03/2003.


\begin{references}
\bibitem{bodzio} Bogdan Damski was very kind to supply the appropriate numerical code.
\end{references}
\end{document}